# *In situ* observation of phase separation and hierarchical microstructure of $K_xFe_{2-y}Se_2$ single crystals

Yong Liu[1*], Qingfeng Xing[1], Warren E. Straszheim[1,2], Jeff Marshman[3], Pal Pedersen[4], and Thomas A. Lograsso[1,5]

[1]*Division of Materials Sciences and Engineering, Ames Laboratory, US DOE, Ames, Iowa 50011, USA*

[2]*Materials Analysis and Research Laboratory, Iowa State University, Ames, Iowa 50011, USA*

[3]*Carl Zeiss Microscopy, LLC, Ion Microscopy Innovation Center (IMIC), Peabody, Massachusetts 01960, USA*

[4]*Carl Zeiss Microscopy, LLC, Thornwood, New York 10594, USA*

[5]*Department of Materials Science and Engineering, Iowa State University, Ames, Iowa 50011, USA*

As solid-state phase transformations occur, materials exhibit rich microstructures depending on the heat treatment conditions, for example quenching temperatures, heating and cooling rates, and pressure [1,2]. The mechanical [3], magnetic [4], transport and superconducting properties [5] of materials can be significantly changed by tuning microstructures upon heat treatment. The $K_xFe_{2-y}Se_2$ single crystals exhibit an iron vacancy order-disorder transition at $T_s$~270 °C [6,7]. Below $T_s$ two spatially separated phases, a minority/superconducting phase and a majority/iron vacancy-ordered phase, were observed through x-ray diffraction [8], transmission electron microscopy (TEM) [9], scanning electron microscopy (SEM) [5,10]. It was found that the superconductivity in the $K_xFe_{2-y}Se_2$ single crystals is strongly dependent on the heat treatment conditions [5]. However, the nature of this phase separation is not well understood. Here, temperature dependent SEM images provide compelling evidence that phase separation corresponds to a nucleation and growth process rather than a spinodal decomposition. The superconducting phase is the remnant of high temperature phase after iron vacancy order-disorder transition. Three-dimension (3D) spatial distribution of superconducting phase in the iron vacancy ordered matrix was revealed by using focused ion beam scanning electron microscopy (FIB-SEM). Our results clearly show that superconducting phase forms a hierarchical structure. Understanding the formation of this hierarchical structure not only can guide the synthesis of bulk superconductors in the future, but also greatly enrich our knowledge about the interplay between phase separation and solid-state phase transformations.

---

[*] Corresponding author: yliu@ameslab.gov

In contrast to other iron based superconductors, $K_xFe_{2-y}Se_2$ compounds [11] are characteristic of phase separation behavior, and they are not bulk superconductors. The coexistence of superconductivity and antiferromagnetic (AFM) order and the interplay between them have been hot topic [12,13]. No hole pockets were observed near the Brillouin zone center in the angle-resolved photoemission spectroscopy studies [14,15], which distinguishes $K_xFe_{2-y}Se_2$ compounds from other iron based superconductors. Different pairing symmetry, either *d* wave [16] or *s* wave [17] seems to be responsible for the superconductivity in these compounds. Although the physics is attractive in $K_xFe_{2-y}Se_2$ compounds, the origin of superconducting phase in $K_xFe_{2-y}Se_2$ compounds remains unclear. One puzzle in the current research imminently needs to be solved: Whether the pure superconducting phase can be synthesized or not. To answer this question, we must clarify how the superconducting phase forms in the $K_xFe_{2-y}Se_2$ single crystals.

Figure 1 shows the temperature dependence of SEM images obtained in the (001) plane. At room temperature, a modulated texture with a bright contrast is visible as the network. This network is weaved by stripes perpendicular to each other. Each individual stripe consists of regularly ordered plates. The crystallographic *a* axis of dark matrix (iron vacancy ordered phase) was determined by back Laue x-ray diffraction studies, which reveals that the strips rotate 45° from *a* axis. The strips orient along the [110] and [1$\bar{1}$0] directions of dark matrix, consistent with previous reports [10,18]. As we already proved, the $K_xFe_{2-y}Se_2$ single crystals quenched above the iron vacancy order-disorder transition temperature $T_s$ have the contiguous network throughout the matrix showing large shielding fraction, whereas furnace-cooling obtained crystals have isolated long and thick strips but very weak shielding fraction [5]. Here, the usage of as-grown crystals or those obtained by quenching treatment does not affect the discussion about the evolution of surface morphology with varying temperatures.

With an increase of temperature approaching $T_s$, we see that strips grow thick and small plates merge into large rectangular bars. The area with white contrast increases and isolated stripes spread and merge into larger ones. The contrast between white strips and dark matrix becomes weak. The temperature dependence of surface morphology clearly reveals that white phase increases its fraction at the expense of the matrix, connects with each other and ultimately the sample becomes single phase above 305 °C. That the contrast between two phases in the SEM images completely disappears suggests the homogeneous solid solution. There are no endothermic or exothermic reactions observed at higher temperatures by examining several of the heating and cooling cycles until the crystals melt incongruently at 900 °C through the differential scanning calorimeter (DSC) measurements [5]. Upon cooling, the SEM images show the visible contrast at 290 °C. At 285 °C, the dark rectangular areas appear. The dark areas further increase while white areas seem to shrink and form checkerboard pattern at 280 °C. The checkerboard pattern does not change too much with a further decrease of temperature. But longitude and latitude lines continuously shrink and become thin. Finally, the white phase aggregates into disconnected strings of rectangular bars. These strips weaves the square network with a side length of ~5μm throughout the (001) plane. In the quenched samples the side length of the network cell can reach 1 μm [5]. Compositions of white phase show slight but detectable change upon heating and cooling, see Table I. As can be seen, potassium content increases while iron content decreases in white phase at high temperatures. Interestingly, dark matrix does not change too much.

TABLE I. Compositions of white phase and dark matrix at different temperatures.

| T (°C) | white phase | dark matrix |
|---|---|---|
| 25 (heating) | $K_{0.54}Fe_{1.8}Se_2$ | $K_{0.75}Fe_{1.53}Se_2$ |
| 254 (cooling) | $K_{0.67}Fe_{1.67}Se_2$ | $K_{0.76}Fe_{1.57}Se_2$ |
| 234 (cooling) | $K_{0.62}Fe_{1.70}Se2$ | $K_{0.76}Fe_{1.56}Se_2$ |
| 214 (cooling) | $K_{0.63}Fe_{1.70}Se_2$ | $K_{0.75}Fe_{1.57}Se_2$ |

To discuss the morphology evolution with varying temperature, we need recall previous structure analysis on the iron vacancy order–disorder transition in $A_xFe_{2-y}Se_2$ compounds (A=alkali elements and Tl) [6,7,19,20]. Above iron vacancy order–disorder transition temperature $T_s$, $K_xFe_{2-y}Se_2$ has a $ThCr_2Si_2$-type tetragonal structure with I4/mmm space group, see Figure 2(a). Both K and Fe sites are randomly partially occupied. It should be pointed out $K_xFe_{2-y}Se_2$ single crystals can host a small amount of iron impurity atoms during crystal growth [5]. The average iron content in $K_xFe_{2-y}Se_2$ single crystals is beyond the ideal iron vacancy ordered phase $K_{0.8}Fe_{1.6}Se_2$. Below $T_s$, the iron vacancy ordered phase has a $\sqrt{5} \times \sqrt{5} \times 1$ superstructure with I4/m space group, as shown in Figure 2(b). Figure 2(c) shows top view (along $c$ axis) of a $3 \times 3 \times 1$ supercell of the iron vacancy ordered phase. It should be emphasized that the dashed cell indicates the $ThCr_2Si_2$-type subcell in the iron vacancy ordered structure. It is very important to identify the composition and crystal structure of superconducting phase in $K_xFe_{2-y}Se_2$ single crystals. Unfortunately, there is no unambiguous answer to this "simple" question. The scanning tunneling microscope (STM) measurements demonstrated that the superconducting phase has a complete FeSe layer or with Se vacancy defects, close to the ideal $ThCr_2Si_2$ structure [21,22]. Nuclear magnetic resonance (NMR) experiments also suggested the superconducting phase in $Rb_{0.74}Fe_{1.6}Se_2$ single crystal to be $Rb_{0.3(1)}Fe_2Se_2$ [23]. Rietveld refinement to high-resolution synchrotron and single-crystal x-ray diffraction data revealed that a metallic superconducting phase $K_xFe_2Se_2$ with x ranging from 0.38 to 0.58 is only observed in large crystals of $K_xFe_{2-y}Se_2$ grown from the melt [24]. Other STM measurements suggested that the superconducting phase consists of a single Fe vacancy for every eight Fe-sites arranged in a $\sqrt{8} \times \sqrt{10}$ parallelogram structure [25]. It should be pointed out that this structure was constructed with a basis of the composition of superconducting phase is about $K_{0.64}Fe_{1.78}Se_2$ ($K_2Fe_7Se_8$ phase) by using energy-dispersive X-ray spectroscope (EDS) built in SEM. The refinements of neutron diffraction data suggest that superconducting phase has a composition of $K_{0.53}Fe_2Se_2$, while the $\sqrt{8} \times \sqrt{10}$ parallelogram structure with the single Fe vacancy does not fit spectra well [20]. It is very likely that the superconducting phase has $ThCr_2Si_2$ structure with complete FeSe layers.

As can be seen in Figure 1, the area of white phase (superconducting phase) in SEM images enlarges upon warming and shrinks upon cooling during the iron vacancy order-disorder transition. Strikingly, our result implies that superconducting phase fraction actually is more prevalent as the temperature is increased, assuming the composition also changes. And the matrix disappears above $T_s$. With a decrease of temperature, the matrix grows up in the superconducting phase. This finding is strongly against our intuition, i.e., superconducting phase precipitates during cooling. With increasing temperature, the matrix tends to transform from iron vacancy ordered state to disordered state. Meanwhile, excess iron atoms in the superconducting phase will diffuse into area with low density of iron atoms (matrix). Remember that iron occupied sites and iron vacancy sites are identical at high temperatures. Therefore, the superconducting phase appears to grow up and phase fraction increases. As temperature is cooled down across $T_s$, iron vacancy ordered phase ($K_{0.8}Fe_{1.6}Se_2$) nucleates and grows up. The phase fraction of dark matrix becomes large and that of white phase reduces. It is matrix that that nucleates or precipitates during cooling. Strips (superconducting phase) are remnant of the high-temperature parent phase with same crystal structure.

For a spinodal decomposition, supersaturated solid solution under the miscibility gap will decompose into two phases through a compositional wave without classic nucleation and growth process [1,2]. Although the regular patterns observed in $K_xFe_{2-y}Se_2$ single crystals are similar to those observed in the compounds showing spinodal decomposition [26], as well as very close crystal structures and coherent structure of two phases, the initial stage is crucial to judge whether spinodal decomposition or normal nucleation and growth happened to the sample. If spinodal decomposition does happen, the boundaries of two phases should always keep, either upon warming or cooling. The driving force for the spinodal decomposition results from the net decrease in free energy of two phases even with small composition fluctuations [1,2]. For $K_xFe_{2-y}Se_2$ single crystals, the iron vacancy order-disorder transition is the driving force of phase separation. The spinodal decomposition has a feature of uphill diffusion [1,2]. As shown in Figure 2(c), the unit cell of $K_{0.8}Fe_{1.6}Se_2$ phase contains a subcell of $KFe_2Se_2$ phase. As the excess iron atoms are expelled from iron vacancy ordered area, $KFe_2Se_2$ subcells will be connected and form iron rich domains. So the formation of iron rich domains should not consume too much energy with the help of iron vacancy order-disorder transition. The size of iron rich domains strongly depends on the completeness of iron vacancy order-disorder transition. Iron rich domains should become more compact at low temperatures. The final strip pattern of iron rich phase is determined by the equilibrium between two phases with considering the degree of iron vacancy ordering, the diffusibility and segregation of iron vacancies. Our results give an unambiguous conclusion that phase separation in $K_xFe_{2-y}Se_2$ single crystals is related to the nucleation and growth process rather than spinodal decomposition. We are lucky to observe the superconductivity in $K_xFe_{2-y}Se_2$ single crystals because the superconducting phase is the remnant of the high temperature phase enduring the iron vacancy order-disorder transition.

Recently, Wang et al. observed the superconductivity in the $K_{2-x}Fe_{4+y}Se_5$ polycrystalline samples annealed at 750 °C and quenched into ice water [27]. No second phase was found in the samples through x-ray diffraction within the instrument resolution, suggesting the absence of phase separation in their superconducting polycrystalline samples. The $\sqrt{5} \times \sqrt{5} \times 1$ superstructure peaks disappeared in the samples. Therefore, Wang et al. suggested that superconductivity in the $K_{2-x}Fe_{4+y}Se_5$ polycrystalline samples result from the frustrated iron vacancy order-disorder transition, where the samples should be in iron vacancy disordered state by quenching [27]. And Iron vacancy ordered $K_{2-x}Fe_{4+y}Se_5$ is the insulating parent compound of the superconducting state [27]. First of all, Wang et al. demonstrated the "bulk superconductivity" in their samples with magnetic susceptibility data. We already proved that large shielding fraction in the magnetization measurements does not mean bulk superconductivity in $K_xFe_{2-y}Se_2$ single crystals [5]. It is necessary to apply specific heat measurement to check whether the samples are bulk superconductors or not. The structure analysis and relative discussions make sense only if $K_{2-x}Fe_{4+y}Se_5$ polycrystalline samples are bulk superconductors. Secondly, the point in Wang et al.'s study is the absence of iron vacancy ordered phase (no phase separation) in the superconducting samples. It should be pointed out that there is no superconductivity observed in polycrystalline powder samples when changing the stoichiometry in $K_xFe_{2-y}Se_2$ where 0.5 < *x* < 1 and 1.4 < *y* < 2 in an early report [24]. Wang et al. observed the superconductivity in the samples annealed at 750 °C, but not in the samples annealed at 300 °C [27]. The heat treatment is of vital necessity in introducing superconductivity in Wang et al.'s samples, which implies that heat treatment plays a role similar to those observed in the single crystals [5]. Their polycrystalline samples were sintered at 700 °C for 24 hours, but the samples were annealed at 750 °C for 1~6 hours, which means that the samples were actually sintered at higher temperature later and grains can grow large.

They observed that superconducting samples contain larger layer-type grains, whereas granular-type grains exist in the non-superconducting samples [27]. SEM images in Wang et al.'s report clearly show that the size of grains in the samples annealed at 300 °C is around 1 μm, while layer-type grains are around serval micrometers. Wang et al. explained that larger layer-type grains are formed because vacant sites are gradually filled in the superconducting sample when the vacancies become more disordered so that the material becomes more two-dimensional [27]. Again the shape and size of grains in the polycrystalline samples rely on the sintering temperatures. But the quenching treatment does help to freeze the iron vacancy disordered state. It is possible that the high-temperature iron vacancy disordered state can be well kept by quenching at 750 °C in Wang et al.'s polycrystalline samples. The quenching at 300 °C does not effectively suppress the iron vacancy order-disorder transition. We believe that high-temperature iron vacancy disordered phase should not be superconducting phase. One should not get bulk superconductor even if the high-temperature iron vacancy disordered state can be frozen by rapid quenching. No matter what quenching temperature was applied, we did not observe bulk superconductivity in the $K_xFe_{2-y}Se_2$ single crystals [5]. Obviously, it is difficult to freeze the iron vacancy disordered state in a large piece of single crystal by the same quenching technique (such as ice water or liquid nitrogen), due to less imperfection and defects in the samples. Even though the iron vacancy order-disorder transition is suppressed in the polycrystalline samples quenched at 750 °C, those samples sould not have bulk superconductivity. It is very likely that the regularly ordered network observed in the single crystals is replaced by the randomly distributed strips in the polycrystalline samples. In that case, x-ray diffraction is not eligible to distinguish the two phases. But local structure detection tools still can discern the two iron environment in the polycrystalline samples. Here, our results provide a reasonable explanation for the highly divergent results in the previous report on the processing techniques and stoichiometry in the starting materials.

SEM images shown in Figure 1 only provide the information about the two-dimensional (2D) spatial distribution of the superconducting phase in (001) plane of majority/iron vacancy ordered phase. Three-dimensional (3D) spatial distribution of the superconducting phase in the matrix was reconstructed through a series of SEM micrographs obtained by successively cross sectioning with focused ion beam (FIB). Figure 3 displays how cross sectional view was obtained by focused ion beam-scanning electron microscopy (FIB-SEM). The cross section was obtained by FIB milling, which was applied parallel to the *c*-axis of crystals. Following each milling sequence, a SEM image is recorded for the exposed cross section. Obviously, the accurateness of reconstructed 3D structure mainly depends on the FIB slice thickness. In this study, the thickness of the slices is 5 nm. With the successive milling and imaging process, a series of SEM images were obtained. The recorded images were stacked into a dataset. Handling and visualization of three-dimensional datasets is straightforward with image analysis software, ORS Visual SI Advanced. A volume of 16×10×10 $\mu m^3$ (16×10 $\mu m^2$ for the cross section and 10 $\mu m$ along slicing direction) was reconstructed with the data treatment. Because the small rectangular plates are of micrometer scale and the size of network cell generally varies from 1 to 4 $\mu m$, the microstructure extracted from the reconstructed volume is reliable with such slice thickness. Figure 3 also shows examples of the cross section of (100) and (010) planes of the $K_xFe_{2-y}Se_2$ single crystals after FIB milling.

The 3D spatial distribution of superconducting phase was shown in the movies as supplementary materials of this report. The microstructure of $K_xFe_{2-y}Se_2$ single crystals consists of interlacing branches. In Figures 4(a)-4(f), the snapshots reveal the typical features of 3D network. As we observe along *c* axis of matrix, the snapshots exhibit a square bowl structure. The four tilted planes of the square bowl consist of strips. The previously observed square network actually consists of the cross section as the strips were cut along *ab* plane. We do not find that any strips orient in the *ab* plane. When we observe along *a* or *b* axis of matrix, the microstructure shows a parallelogram shape. The interweaved strips seem to intersect and show a X shape, as shown in Figures 4(d) and 4(e). But most of the strips do not intersect in the space if we rotate the 3D microstructure. Let us follow the trace of strips. We find that the strips can be grouped into different planes. The strips grow parallel to each other in this plane. The strips grow straightly in this plane with limited length, almost same scale with the network cell. Then the strips twist, bend, furcate and grow in a new plane. Indirect measurements such as NMR [23], muon spin rotation (μSR) [28], and Mössbauer [29] spectroscopy experiments have revealed nearly 90% of the sample volumes exhibit large-moment antiferromagnetic (AFM) order, while 10% of the sample volumes remain paramagnetic (PM) and attributed to a metallic/superconducting phase in $A_xFe_{2-y}Se_2$ single crystals. In this study, we directly obtain that the volume fraction of superconducting phase is between 10% and 12%, which well matches the previous results.

In a previous report, it was suggested that the microstructure of superconducting phase consists of hollow truncated octahedron similar with what discussed for Archimedean solids [30]. It should be pointed out that the hypothesis of Archimedean solidlike superconducting framework was based on the strip patterns in the (001), (100), and (110) planes. Our direct observation of microstructure of $K_xFe_{2-y}Se_2$ single crystals reveals that the strips actually wrap the matrix and form a parallelepiped, as shown in Figure 4(g). By simply cutting the $K_xFe_{2-y}Se_2$ single crystals along (100) plane, SEM image shown in Figure 4(h) confirms that $K_xFe_{2-y}Se_2$ single crystals are made of closely packed parallelepipeds with a size of (1-2)×(3-4) $\mu m^2$. The observed strip patterns in the SEM images of (001), (100) and (010) planes shown in Figure 3 correspond to the cross sections of the strips in the planes of the parallelepiped. There exist two geometry relationships in the microstructure of $K_xFe_{2-y}Se_2$ single crystals: the growth habit of superconducting phase and orientations of strips in the dark matrix. It should be pointed out that the superconducting phase was embedded in the matrix. There are no facets exposed to the viewer. We only can observe the cross sections of superconducting phase in different FIB milled planes. As shown in Figures 3 and 4(i), the angle between the intersection line and *c* axis was measured about 30°. Therefore, Speller et al. suggested that the strips observed in (100) plane of matrix correspond to {113} habit planes [10,31]. Figure 2(a) illustrates the geometry relationship between (113) plane and *c* axis of superconducting phase. It should be pointed out that such geometry relationship should not be applied to judge the facet as (113) plane because what we know is the intersection line between the facets of superconducting phase and (001) or (100) /(010) planes. But these facets can be tilted and they are not perpendicular to (100) plane. The determination of growth habit needs to be done by the combination of electron backscatter diffraction (EBSD) and grain boundary trace analysis. As for the orientation of strips relative to the matrix, Figure 2(b) displays that the intersection line between (111) and (001) planes orients along [110] direction, and The angle between the c axis and intersection line between (111) and (100) planes is about 30°. We can determine that strips are located at the {111} planes of the matrix. The (111) plane of matrix was compared with (113) plane of superconducting phase suggested by Speller et al [10,31] in Figure 2(d). It is suggested that that planes of parallelepiped correspond to the {111} planes of matrix. It is very likely that strips grow along [111] direction.

    This growth direction may be due to anisotropy in the elastic energy along *c* axis and in the *ab* plane, obviously related the crystallographic symmetry [32]. The microstructural length scale is determined by the interplay or balance between the elastic energy and vacancy ordering. The superconducting phase has a smaller *a* axis parameter but larger *c* axis parameter compared to

the iron vacancy ordered phase [20,24]. The *a* axis parameter of iron vacancy ordered phase needs to be transferred with the relation $a = \sqrt{5}a'$. It is suggested that the tilted strips should minimize the elastic energy [32]. The superconducting domains tend to be compressed in the *ab* plane by iron vacancy ordered phase, as shown in Figure 4(j). Meanwhile the superconducting domains expand along the direction that has an angle 30° from *c* axis, as shown in Figure 4(k). With an increase of length of strips, the elastic strain accumulates along *c* axis. Now, we can understand that the edges of parallelepiped are thick strips because they are suppressed by the four adjacent parallelepipeds. The thin strips are observed in the plane connecting two parallelepipeds because of less pressure. Strips have limited length because the elastic energy must be released. Then Strips grow along another direction. Here, the superconducting phase in the matrix constructs new structure, which is named as hierarchical structure [33]. This structural hierarchy can play a key role in determining the bulk material properties.

For the phase separation in the $K_xFe_{2-y}Se_2$ single crystals, solving the problem how the superconducting phase forms has the highest priority. Our study provides strong evidence that the superconducting phase is the remnant of high temperature phase after iron vacancy order-disorder transition. *The superconducting phase should not form without iron vacancy order-disorder transition.* It is impossible to separate the superconducting phase from the matrix. The hierarchical microstructure of $K_xFe_{2-y}Se_2$ single crystals was observed by using FIB-SEM, which should result from the balance between elastic energy and iron vacancy ordering.

**Methods**

To grow $K_xFe_{2-y}Se_2$ single crystals, the nominal $K_{0.8}Fe_{2+z}Se_2$ polycrystalline samples were synthesized with more iron (*z*=0, 0.2, and 0.6) adding in the starting materials. The powder was sealed in a small quartz ampoule with a diameter of 0.8 cm. This small ampoule was then sealed in a bigger ampoule. The crystal growth was performed in a vertical tube furnace. The ampoule was heated to 1030 °C and held for 2 hours. The furnace was cooled down to 780 °C at a cooling rate of 6 °C/h. see more details about heating conditions and heat treatment in Ref. [5].

The microstructure of $K_xFe_{2-y}Se_2$ single crystals was analyzed with FEI Quanta-250 scanning electron microscopy using backscattered electrons (BSE) and secondary electrons (SE). The actual compositions of crystals were determined by an Oxford energy-dispersive x-ray analysis system.

The 3D microstructure was produced by focused Ion Beam Scanning Electron Microscopy (FIB-SEM) using the Carl Zeiss Crossbeam 540 system. The data was reconstructed by use of ORS Visual Advanced SI 3D Visualization software.

**Acknowledgments**

The work at Ames Laboratory was supported by the U.S. Department of Energy (DOE), Office of Basic Energy Sciences, Materials Science and Engineering Division. Ames Laboratory is operated for the U.S. DOE by Iowa State University under Contract No. DE-AC02-07CH11358.

**Author contributions**

Y.L. conceived the research project and grew the $K_xFe_{2-y}Se_2$ single crystals. W.E.S. conducted the SEM experiments. J.M. conducted FIB-SEM experiments. P.P. conducted the 3D Volume rendering using the ORS software. Y.L. analyzed the data and prepared the manuscript. Y.L., T.A.L. and Q.X. discussed the results. T.A.L. supervised and supported the work.

**Additional information**

Correspondence and requests for materials should be addressed to Y.L. (<yliu@ameslab.gov>).

**Competing financial interests**

The authors declare no competing financial interests.

**References**

1. J. W. Christian, The Theory of Transformations in Metals and Alloys (Part I + II), 3rd edition, Pergamon Press, Oxford (2002).

2. David A. Porter and K. E. Easterling, Phase Transformations in Metals and Alloys Paperback, Chapman & Hall; 2nd edition, London (1992).

3. Florian Vogel, Nelia Wanderka, Zoltan Balogh, Mohammed Ibrahim, Patrick Stender, Guido Schmitz, and John Banhart, Mapping the evolution of hierarchical microstructures in a Ni-based superalloy. Nat. Commun. **4**, 2955 (2013).

4. E. P. Butler and G. Thomas, Structure and properties of spinodally decomposed Cu-Ni-Fe alloys. Acta Metallurgica **18**, 347-365 (1970).

5. Y. Liu, Q. Xing, K. W. Dennis, R. W. McCallum, and T. A. Lograsso, Evolution of precipitate morphology during heat treatment and its implications for the superconductivity in $K_xFe_{1.6+y}Se_2$ single crystals. Phys. Rev. B **86**, 144507 (2012).

6. W. Bao, Q. Z. Huang, G. F. Chen, M. A. Green, D. M. Wang, J. B. He, and Y. M. Qiu, A Novel Large Moment Antiferromagnetic Order in $K_{0.8}Fe_{1.6}Se_2$ Superconductor. Chin. Phys. Lett. **28**, 086104 (2011).

7. A. Ricci, N. Poccia, B. Joseph, G. Arrighetti, L. Barba, J. Plaisier, G. Campi, Y. Mizuguchi, H. Takeya, Y. Takano, N. L. Saini and A. Bianconi, Intrinsic phase separation in superconducting $K_{0.8}Fe_{1.6}Se_2$ ($T_c$ = 31.8 K) single crystals. Supercond. Sci. Technol. **24**, 082002 (2011).

8. A. Ricci, N. Poccia, G. Campi, B. Joseph, G. Arrighetti, L. Barba, M. Reynolds, M. Burghammer, H. Takeya, Y. Mizuguchi, Y. Takano, M. Colapietro, N. L. Saini, and A. Bianconi, Nanoscale phase separation in the iron chalcogenide superconductor $K_{0.8}Fe_{1.6}Se_2$ as seen via scanning nanofocused x-ray diffraction. Phys. Rev. B **84**, 060511(R).

9. Z. Wang, Y. J. Song, H. L. Shi, Z. W. Wang, Z. Chen, H. F. Tian, G. F. Chen, J. G. Guo, H. X. Yang, and J. Q. Li, Microstructure and ordering of iron vacancies in the superconductor system $K_yFe_xSe_2$ as seen via transmission electron microscopy. Phys. Rev. B **83**, 140505(R) (2011).

10. S. C. Speller, T. B. Britton, G. M. Hughes, A. Krzton-Maziopa, E. Pomjakushina, K. Conder, A. T. Boothroyd, C. R. M. Grovenor, Microstructural analysis of phase separation in iron chalcogenide superconductors. Supercond. Sci. Technol. **25**, 084023 (2012).

11. J. G. Guo, S. F. Jin, G. Wang, S. C. Wang, K. X. Zhu, T. T. Zhou, M. He, and X. L. Chen, Superconductivity in the iron selenide $K_xFe_2Se_2$ (0≤x≤1.0). Phys. Rev. B **82**, 180520(R) (2010).

12. E. Dagotto, The unexpected properties of alkali metal iron selenide superconductors. Rev. Mod. Phys. **85**, 849 (2013).

13. W. Bao, Structure, magnetic order and excitations in the 245 family of Fe-based superconductors. J. Phys.: Condens. Matter **27**, 023201 (2015).

14. T. Qian, X.-P. Wang, W.-C. Jin, P. Zhang, P. Richard, G. Xu, X. Dai, Z. Fang, J.-G. Guo, X.-L. Chen, and H. Ding, Absence of a Holelike Fermi Surface for the Iron-Based $K_{0.8}Fe_{1.7}Se_2$ Superconductor Revealed by Angle-Resolved Photoemission Spectroscopy. Phys. Rev. Lett. **106**, 187001 (2011).

15. Y. Zhang, L. X. Yang, M. Xu, Z. R. Ye, F. Chen, C. He, H. C. Xu, J. Jiang, B. P. Xie, J. J. Ying, X. F. Wang, X. H. Chen, J. P. Hu, M. Matsunami, S. Kimura, and D. L. Feng, Nodeless superconducting gap in $A_xFe_2Se_2$ (A=K,Cs) revealed by angle-resolved photoemission spectroscopy. Nature Materials **10**, 273–277 (2011).

16. T. A. Maier, S. Graser, P. J. Hirschfeld, and D. J. Scalapino, *d*-wave pairing from spin fluctuations in the KxFe2−ySe2 superconductors. Phys. Rev. B **83**, 100515(R) (2011).

17. T. Saito, S. Onari, and H. Kontani, Emergence of fully gapped $s_{++}$-wave and nodal *d*-wave states mediated by orbital and spin fluctuations in a ten-orbital model of $KFe_2Se_2$. Phys. Rev. B **83**, 140512(R) (2011).

18. Z.-W. Wang, Z. Wang, Y.-J. Song, C. Ma, Y. Cai, Z. Chen, H.-F. Tian, H.-X. Yang, G.-F. Chen, and J.-Q. Li, Structural Phase Separation in $K_{0.8}Fe_{1.6+x}Se_2$ Superconductors. J. Phys. Chem. C **116**, 17847-17852 (2012).

19. V. Yu. Pomjakushin, D. V. Sheptyakov, E. V. Pomjakushina, A. Krzton-Maziopa, K. Conder, D. Chernyshov, V. Svitlyk, and Z. Shermadini, Iron-vacancy superstructure and possible room-temperature antiferromagnetic order in superconducting $Cs_yFe_{2-x}Se_2$. Phys. Rev. B **83**, 144410 (2011).

20. S. V. Carr, D. Louca, J. Siewenie, Q. Huang, A. Wang, X. Chen, and P. Dai, Structure and composition of the superconducting phase in alkali iron selenide $K_yFe_{1.6+x}Se_2$. Phys. Rev. B **89**, 134509 (2014).

21. W. Li, H. Ding, P. Deng, K. Chang, C. Song, K. He, L. Wang, X. Ma, J.-P. Hu, X. Chen, and Q.-K. Xue, Phase separation and magnetic order in K-doped iron selenide superconductor. Nature Physics **8**, 126–130 (2012).

22. W. Li, H. Ding, Z. Li, P. Deng, K. Chang, K. He, S. Ji, L. Wang, X. Ma, J.-P. Hu, X. Chen, and Q.-K. Xue, $KFe_2Se_2$ is the Parent Compound of K-Doped Iron Selenide Superconductors. Phys. Rev. Lett. **109**, 057003 (2012).

23. Y. Texier, J. Deisenhofer, V. Tsurkan, A. Loidl, D. S. Inosov, G. Friemel, and J. Bobroff, NMR Study in the Iron-Selenide $Rb_{0.74}Fe_{1.6}Se_2$: Determination of the Superconducting Phase as Iron Vacancy-Free $Rb_{0.3}Fe_2Se_2$. Phys. Rev. Lett. **108**, 237002 (2012).

24. D. P. Shoemaker, D. Y. Chung, H. Claus, M. C. Francisco, S. Avci, A. Llobet, and M. G. Kanatzidis, Phase relations in $K_xFe_{2-y}Se_2$ and the structure of superconducting $K_xFe_2Se_2$ via high-resolution synchrotron diffraction. Phys. Rev. B **86**, 184511 (2012).

25. X. Ding, D. Fang, Z. Wang, H. Yang, J. Liu, Q. Deng, G. Ma, C. Meng, Y. Hu, and H.-H. Wen, Influence of microstructure on superconductivity in $K_xFe_{2-y}Se_2$ and evidence for a new parent phase $K_2Fe_7Se_8$. Nature Communications **4**, 1897 (2013).

26. F. Findik, Improvements in spinodal alloys from past to present. Materials and Design **42**, 131–146 (2012).

27. C.-H. Wang, T.-K. Chen, C.-C. Chang, C.-H. Hsu, Y.-C. Lee, M.-J. Wang, P. M. Wu, M.-K. Wu, Disordered Fe vacancies and superconductivity in potassium-intercalated iron selenide ($K_{2-x}Fe_{4+y}Se_5$). arXiv:1502.01116 (2015).

28. Z. Shermadini, H. Luetkens, R. Khasanov, A. Krzton-Maziopa, K. Conder, E. Pomjakushina, and H-H. Klauss, and A. Amato, Superconducting properties of single-crystalline $A_xFe_{2-y}Se_2$ (A=Rb, K) studied using muon spin spectroscopy. Phys. Rev. B **85**, 100501(R) (2012).

29. V. Ksenofontov, G. Wortmann, S. A. Medvedev, V. Tsurkan, J. Deisenhofer, A. Loidl, and C. Felser, Phase separation in superconducting and antiferromagnetic $Rb_{0.8}Fe_{1.6}Se_2$ probed by Mössbauer spectroscopy. Phys. Rev. B **84**, 180508(R) (2011).

30. Z. Wang, Y. Cai, Z. W. Wang, C. Ma, Z. Chen, H. X. Yang, H. F. Tian, and J. Q. Li, Archimedean solidlike superconducting framework in phase-separated $K_{0.8}Fe_{1.6+x}Se_2$ ($0 \leq x \leq 0.15$). Phys. Rev. B **??**, xxxxx (2015). (arXiv:1401.1001)

31. S. C. Speller, P. Dudin, S. Fitzgerald, G. M. Hughes, K. Kruska, T. B. Britton, A. Krzton-Maziopa, E. Pomjakushina, K. Conder, A. Barinov, and C. R. M. Grovenor,

High-resolution characterization of microstructural evolution in Rb$_x$Fe$_{2-y}$Se$_2$ crystals on annealing. Phys. Rev. B **90**, 024520 (2014).

32. M. Doi, Elasticity effects on the microstructure of alloys containing coherent precipitates. Prog. Mater. Sci. **40**, 79–180 (1996).

33. R. Lakes, Materials with structural hierarchy, Nature **361**, 511 - 515 (1993).

**Figure captions**

Figure 1 Temperature dependence of SEM images obtained in the (001) plane of $K_xFe_{2-y}Se_2$ single crystals.

Figure 2 (a) Crystal structure of $KFe_2Se_2$ with $ThCr_2Si_2$-type tetragonal structure (I4/mmm), which is identical to that of high-temperature iron vacancy disordered phase. (b) Crystal structure of $K_{0.8}Fe_{1.6}Se_2$ with ordered iron vacancies (I4/m). (c) A diagramatic representation of (001) plane of the vacancy ordered structure in $K_{0.8}Fe_{1.6}Se_2$. The dashed and dotted lines correspond to the unit cells of $KFe_2Se_2$ and $K_{0.8}Fe_{1.6}Se_2$. The lattice vectors rotate 30° for the two phases. (d) Yellow triangle illustrates the (113) plane in the $KFe_2Se_2$, and blue one represents the (111) plane in $K_{0.8}Fe_{1.6}Se_2$.

Figure 3 Schematic illustration of FIB-SEM serial sectioning procedure, which shows arrangement of FIB and SEM columns in Crossbeam systems with an inclination angle of 54°. The plate-like $K_xFe_{2-y}Se_2$ single crystal was sliced along *a/b* axis of dark matrix. The thickness of the slices is 5 nm. The surfaces of (001), (100) and (010) planes of FIB milled volume were shown.

Figure 4 (a) Video snapshot of 3D reconstructed microstructure of superconducting phase in the $K_xFe_{2-y}Se_2$ single crystals along Z/*c* axis. (b) Video snapshot along milling direction, Y/*b* axis. (c) Snapshot of (001) plane. (d) Snapshot of (100) plane. (e) Snapshot of (010) plane. (f) Corner view of 3D microstructure. Upper half of the picture is (001) plane and the lower half is (100) plane. Figures (c)-(f) shows the interlacing branch structure of superconducting phase in a volume of 8×8×8 μm³ (g) Schematic drawing of a parallelepiped with strips on the two surfaces.

In order to clearly show the results, the patterns in other surfaces are omitted. (h) SEM image of (100) plane, which clearly shows the closely packed parallelepipeds. (i) A thin slice 16×10 μm² of (100) plane. (j) Schematic drawing of phase separation in the (001) plane of $K_xFe_{2-y}Se_2$ single crystals. The superconducting phase has the $ThCr_2Si_2$-type tetragonal structure, where iron vacancy sites (Fe2 sites) are filled with blue balls. The iron vacancy ordered phase has the $\sqrt{5} \times \sqrt{5} \times 1$ superstructure. The arrows indicate the compressive stress on the superconducting phase. Lattice distortions in the interface between two phases are omitted. (k) Phase separation in the (100) plane of $K_xFe_{2-y}Se_2$ single crystals. The *c*-axis lattice parameter of superconducting phase is larger than that of iron vacancy ordered phase. The growth of strips would push away the iron vacancy ordered phase. But they would stop growing along *c* axis as the elastic energy reaches balance between the two phases.

| 31 °C | 250 °C | 270 °C |
|---|---|---|
| 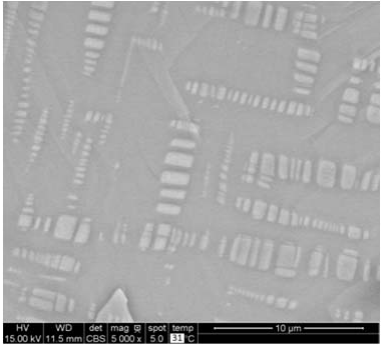 | 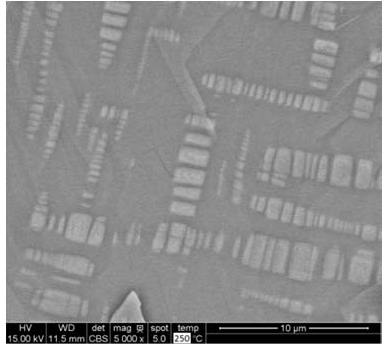 | 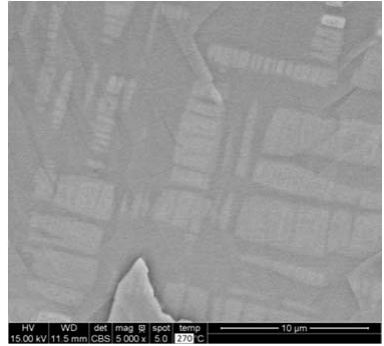 |
| 275 °C | 280 °C | 285 °C |
| 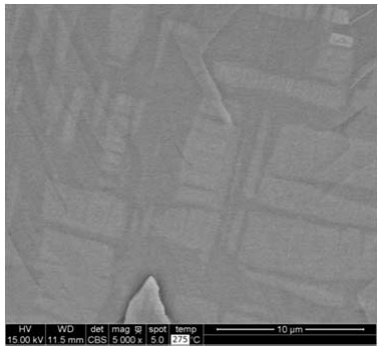 | 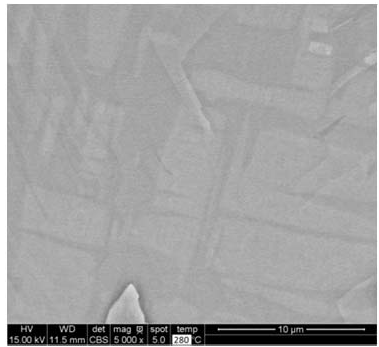 | 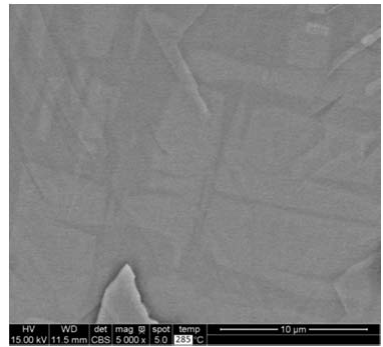 |
| 290 °C | 295 °C | 300 °C |
| 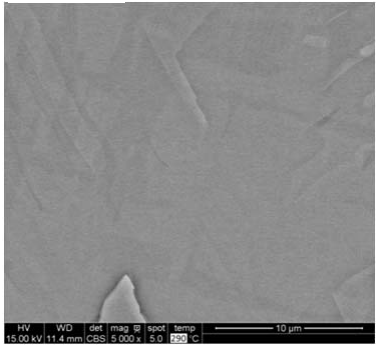 | 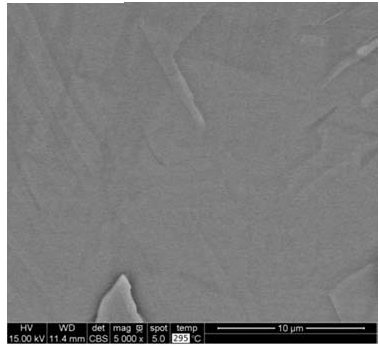 | 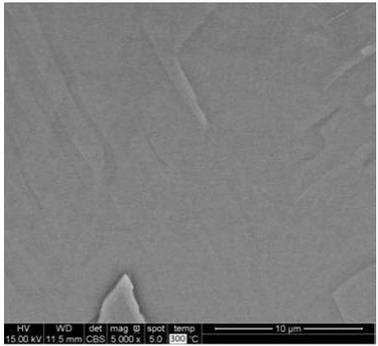 |

305 °C 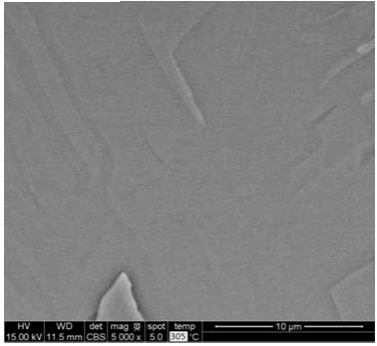 310 °C 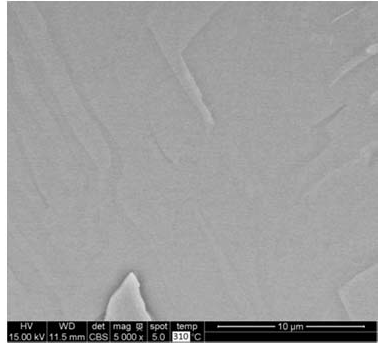 290 °C 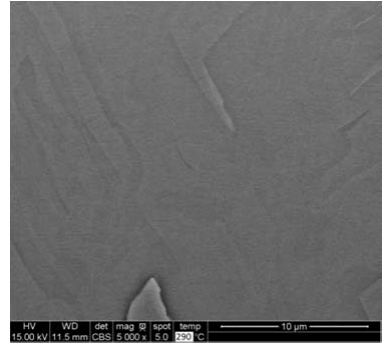

285 °C 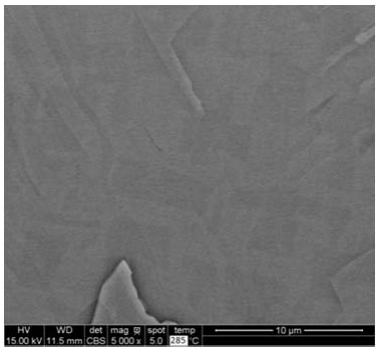 280 °C 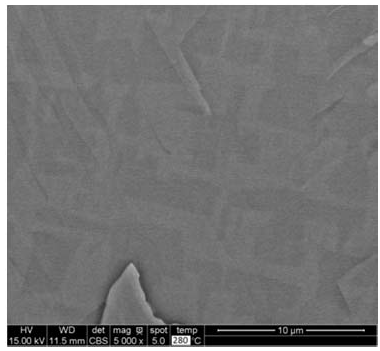 275 °C 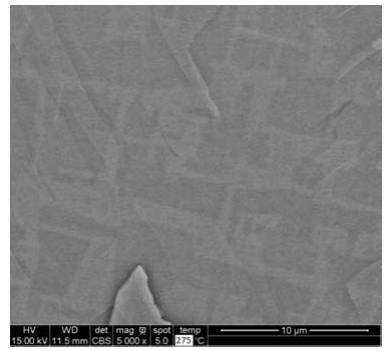

270 °C 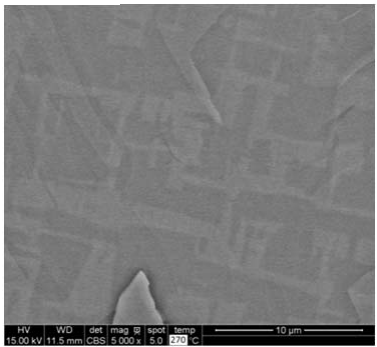 266 °C 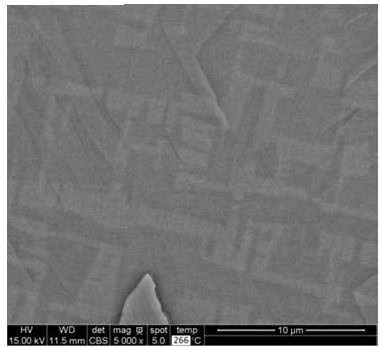 260 °C 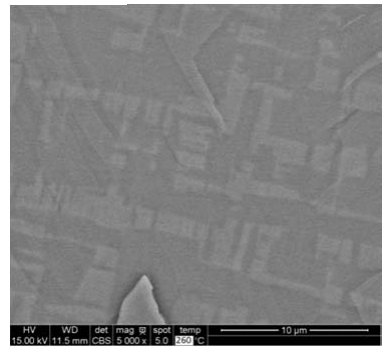

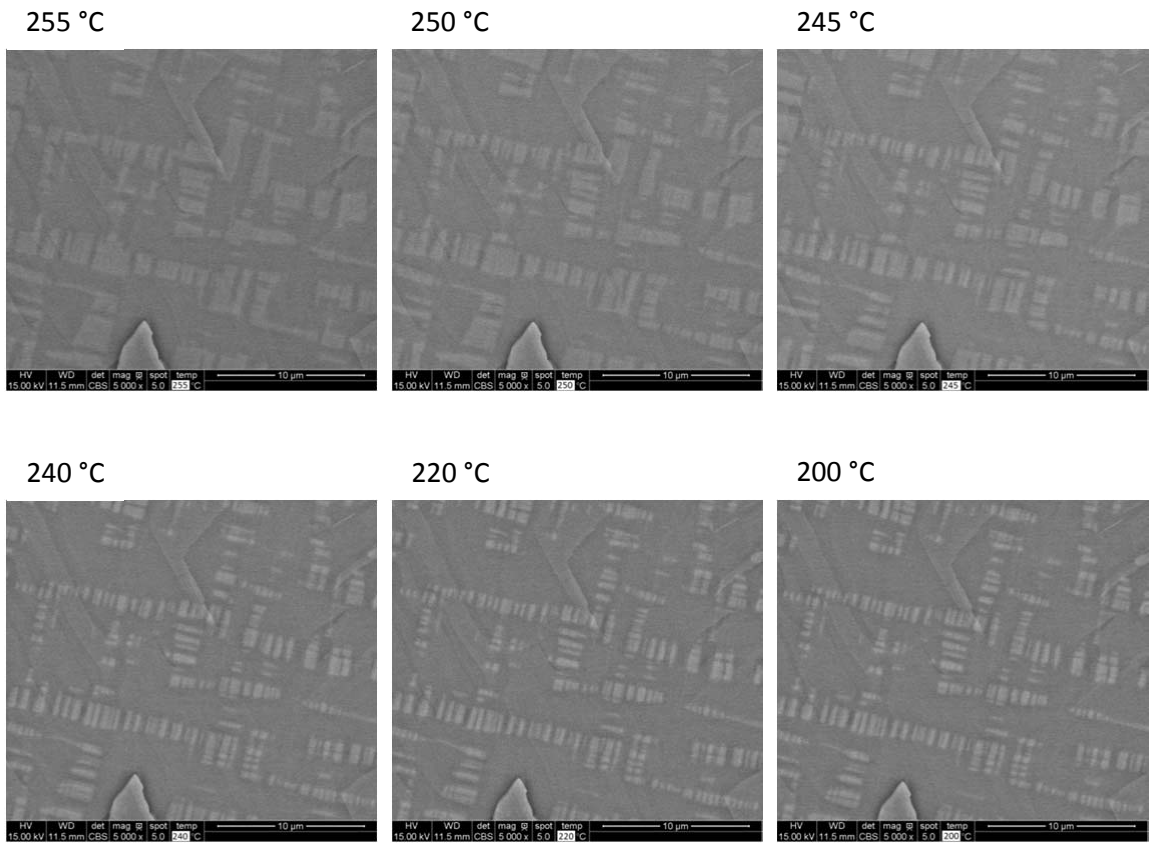

Figure 1

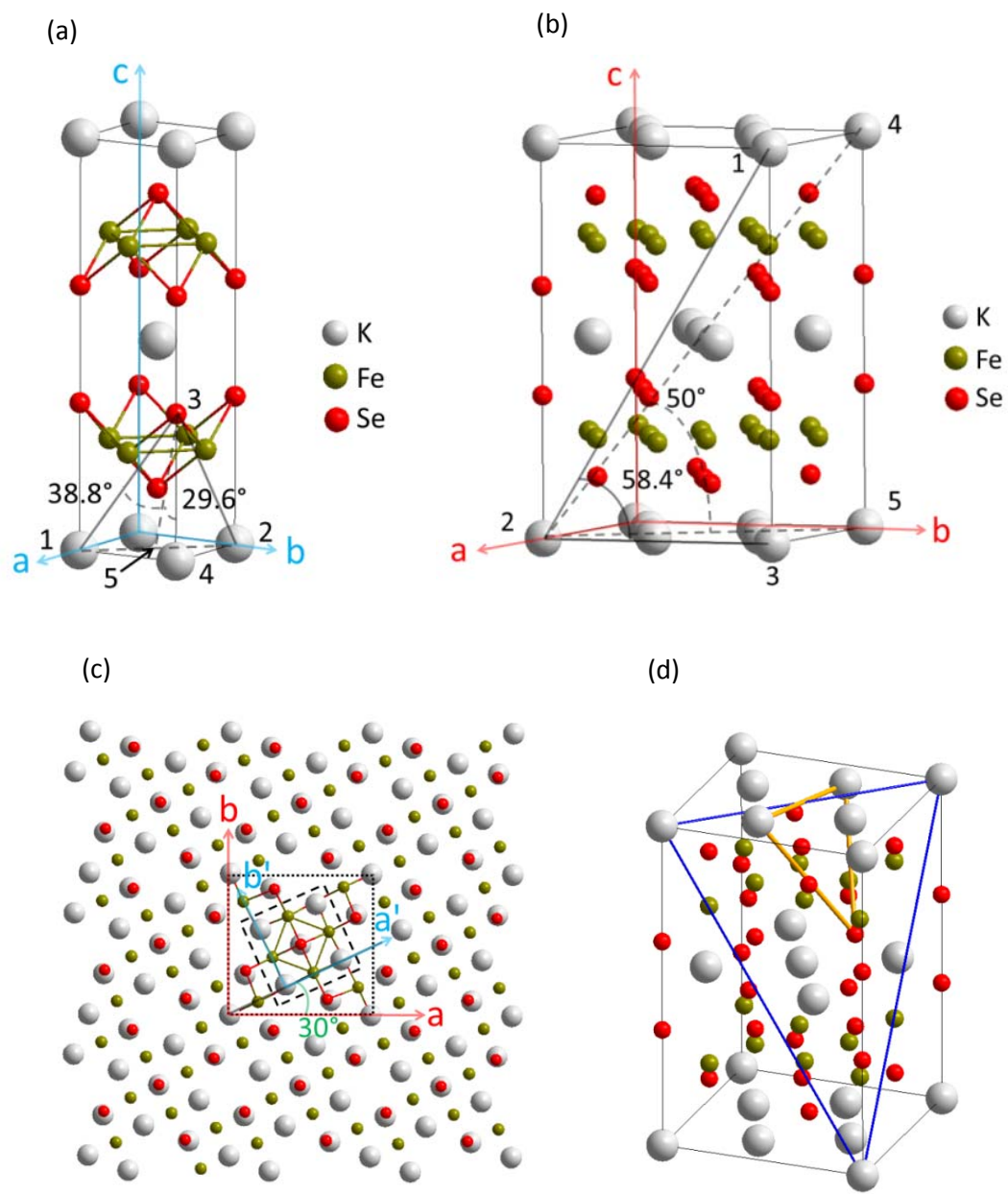

Figure 2

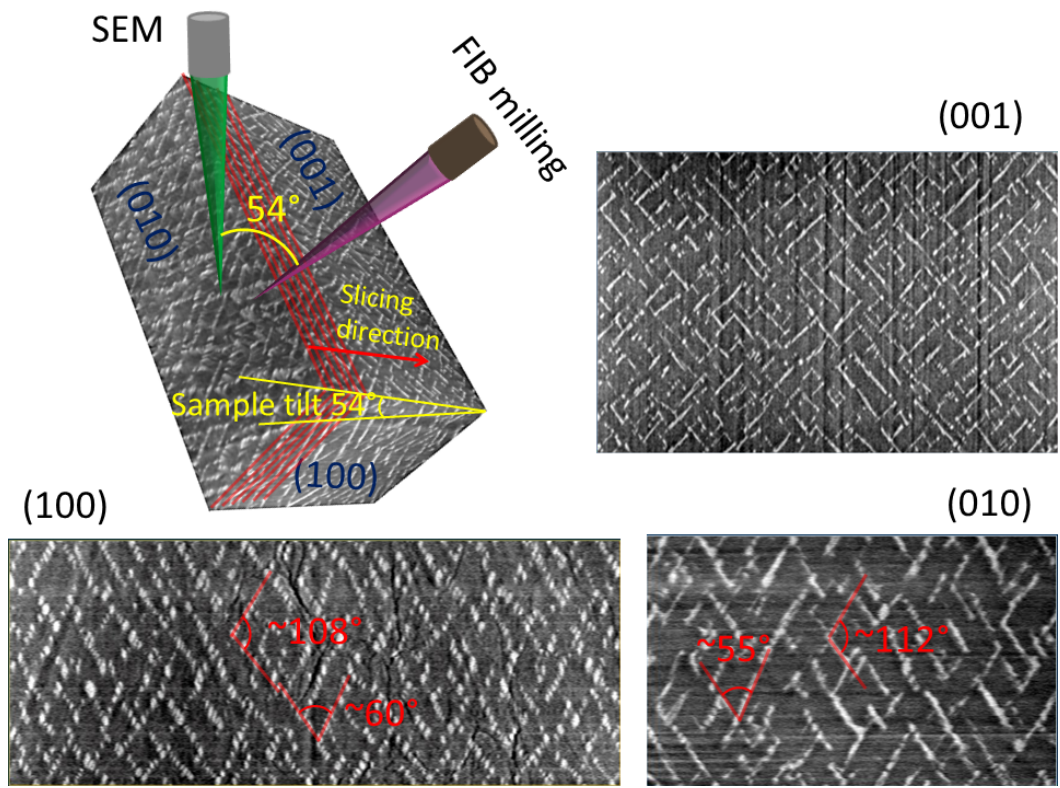

Figure 3

(a)
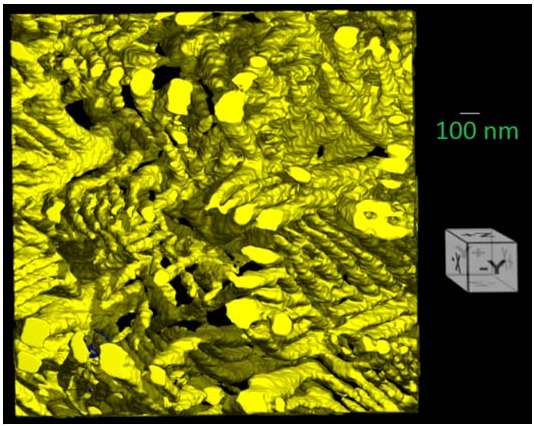

(b)
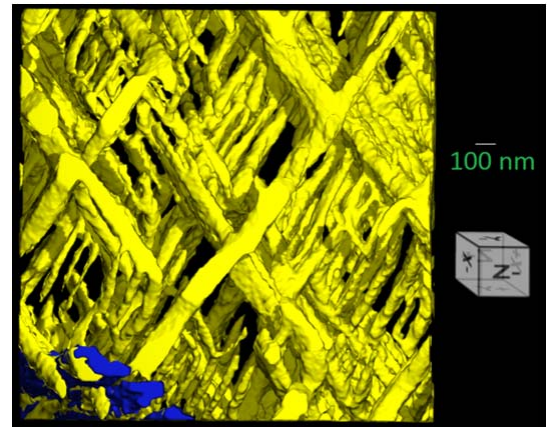

(c)
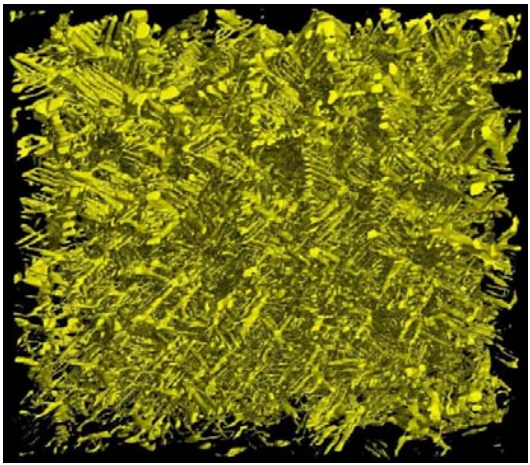

(d)
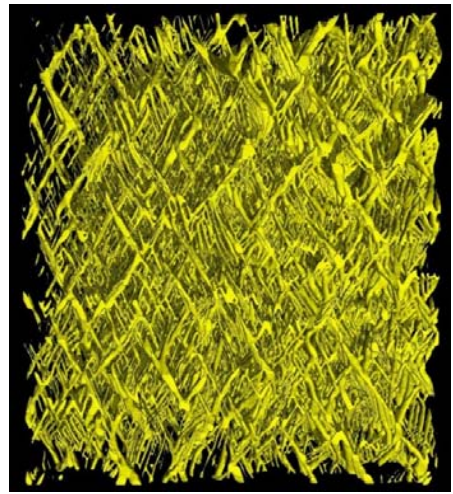

(e)
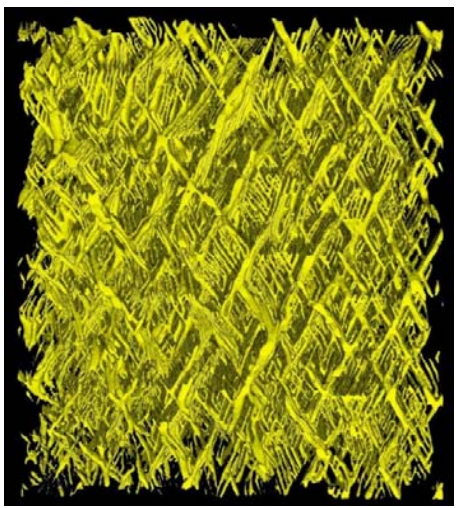

(f)
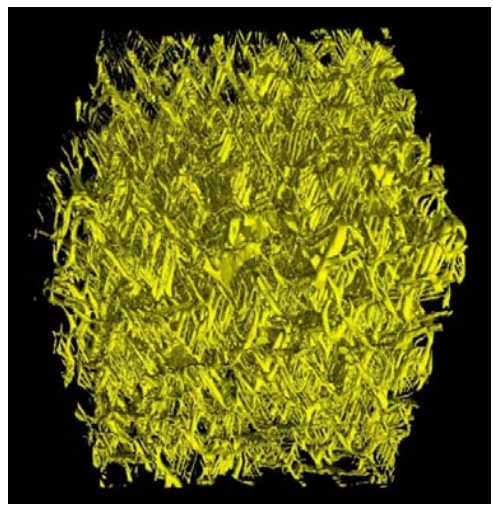

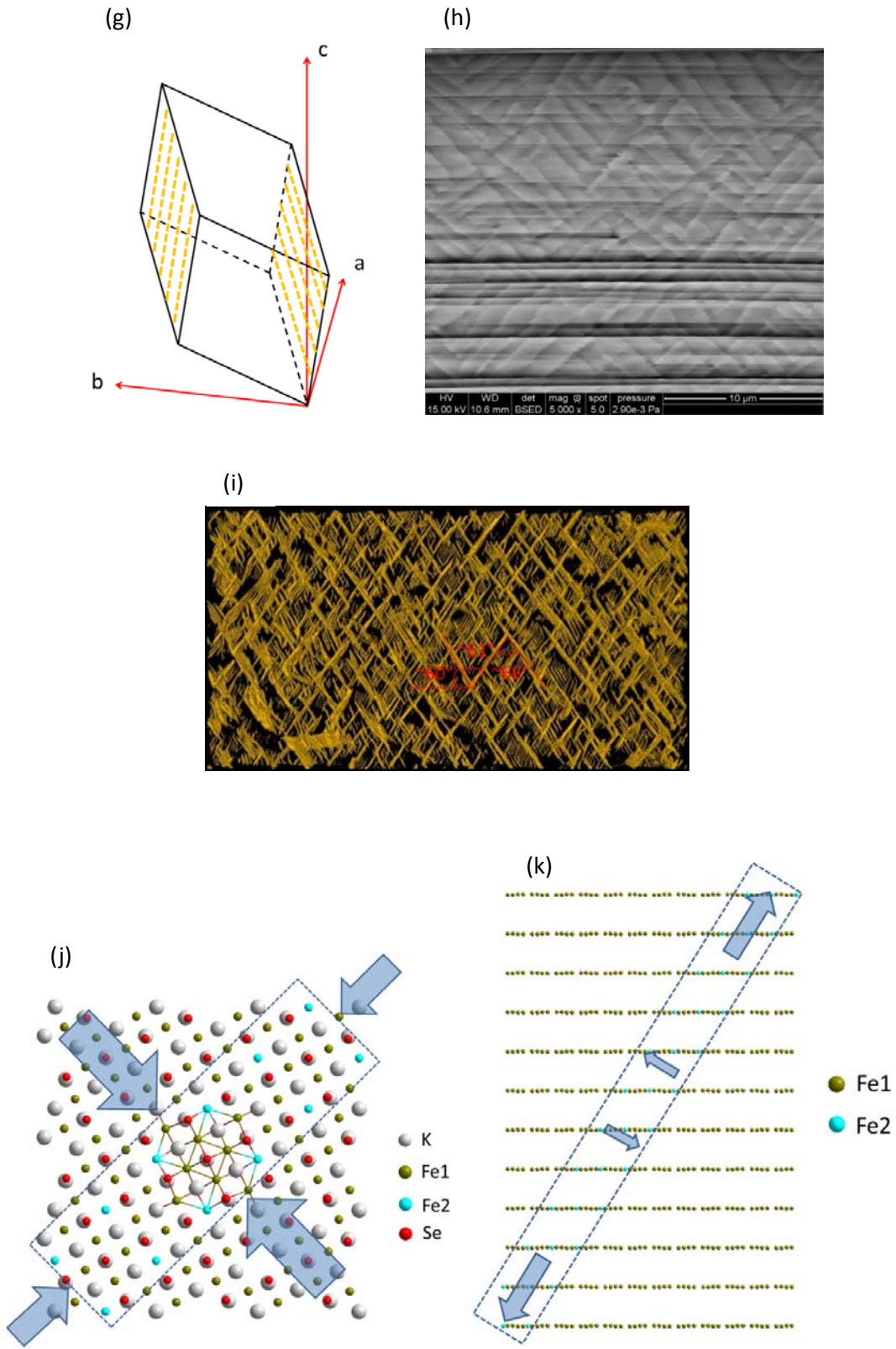

Figure 4